\documentclass[sigconf]{acmart}  

\usepackage{color}
\usepackage{balance}

\def\BibTeX{{\rm B\kern-.05em{\sc i\kern-.025em b}\kern-.08emT\kern-.1667em\lower.7ex\hbox{E}\kern-.125emX}}

\copyrightyear{2024}
\acmYear{2024}
\setcopyright{rightsretained}
\acmConference[CHI EA '24]{Extended Abstracts of the CHI Conference on Human Factors in Computing Systems}{May 11--16, 2024}{Honolulu, HI, USA}
\acmBooktitle{Extended Abstracts of the CHI Conference on Human Factors in Computing Systems (CHI EA '24), May 11--16, 2024, Honolulu, HI, USA}
\acmDOI{10.1145/3613905.3647967}
\acmISBN{979-8-4007-0331-7/24/05}

\begin{document}

%
\title{Enhancing Programming Error Messages in Real Time with Generative AI}


%
\author{Bailey Kimmel}
\orcid{0009-0000-6655-0564}
\affiliation{
  \institution{Abilene Christian University}
  \city{Abilene, Texas}
  \country{United States}
}
\email{blk20c@acu.edu}

\author{Austin Geisert }
\orcid{0009-0000-0058-5254}
\affiliation{
  \institution{Abilene Christian University}
  \city{Abilene, Texas}
  \country{United States}
}
\email{alg22b@acu.edu}

\author{Lily Yaro}
\orcid{0009-0003-4080-3676}
\affiliation{
  \institution{Abilene Christian University}
  \city{Abilene, Texas}
  \country{United States}
}
\email{lcy20a@acu.edu}

\author{Brendan Gipson}
\orcid{0009-0003-3229-2220}
\affiliation{
  \institution{Abilene Christian University}
  \city{Abilene, Texas}
  \country{United States}
}
\email{bmg20b@acu.edu}

\author{Taylor Hotchkiss}
\orcid{0009-0004-1081-2296}
\affiliation{
  \institution{Abilene Christian University}
  \city{Abilene, Texas}
  \country{United States}
}
\email{rth19b@acu.edu}

\author{Sidney Osae-Asante }
\orcid{0009-0006-0732-9038}
\affiliation{
  \institution{Abilene Christian University}
  \city{Abilene, Texas}
  \country{United States}
}
\email{sko21a@acu.edu}

\author{Hunter Vaught}
\orcid{0009-0003-4564-9508}
\affiliation{
  \institution{Abilene Christian University}
  \city{Abilene, Texas}
  \country{United States}
}
\email{hrv19a@acu.edu}

\author{Grant Wininger}
\orcid{0009-0009-0339-0618}
\affiliation{
  \institution{Abilene Christian University}
  \city{Abilene, Texas}
  \country{United States}
}
\email{tgw20a@acu.edu}

\author{Chase Yamaguchi}
\orcid{0009-0007-3198-6318}
\affiliation{
  \institution{Abilene Christian University}
  \city{Abilene, Texas}
  \country{United States}
}
\email{chy20a@acu.edu}

%
\renewcommand{\shortauthors}{B. Kimmel et al.}

%
\begin{abstract}
Generative AI is changing the way that many disciplines are taught, including computer science. Researchers have shown that generative AI tools are capable of solving programming problems, writing extensive blocks of code, and explaining complex code in simple terms. Particular promise has been shown in using generative AI to enhance programming error messages. Both students and instructors have complained for decades that these messages are often cryptic and difficult to understand. Yet recent work has shown that students make fewer repeated errors when enhanced via GPT-4. We extend this work by implementing feedback from ChatGPT for all programs submitted to our automated assessment tool, Athene, providing help for compiler, run-time, and logic errors. Our results indicate that adding generative AI to an automated assessment tool does not necessarily make it better and that design of the interface matters greatly to the usability of the feedback that GPT-4 provided.
\end{abstract}



%
\maketitle

%
%

\section{Introduction}
The introduction of artificial intelligence (AI) in the form of large language models (LLMs) is changing many disciplines, including computer science education \cite{denny2023computing}. Models such as GPT-3 and GPT-4 and tools such as Github Copilot have upended decades of pedagogical wisdom \cite{becker2023programming}. These tools can write, explain, and debug code in ways that simply were not possible just two years ago \cite{prather2023robots}. Researchers have been quick to measure the ability of these LLMs with regard to typical computer science programs \cite{finnieansley2022robots, finnieansley2023my}, code explanations \cite{leinonen2023comparing}, exams \cite{mahon2023no}, and even Parsons Problems (mixed-up code problems) \cite{reeves2023evaluating}. This has led to instructors having many concerns, such as students using autogenerated code that they don't understand \cite{prather2024tochi} or using these tools to do their work for them \cite{lau2023ban, becker2023programming}. Still other concerns, such as over-reliance, biases inherent in the models, and trustworthiness, remain active points of discussion in current research \cite{zastudil2023generative}.

Despite the alarm, generative AI has positive uses. For instructors, these models can automatically generate programming exercises \cite{sarsa2022automatic} and code explanations \cite{macneil2023experiences}. They can even be used to create entirely new types of problems for students to solve, such as Prompt Problems, which ask students to write the prompt that would solve a given problem \cite{denny2024prompt}. For students, generative AI can be a useful tool for learning when it's used to generate code a student already knows how to write \cite{prather2023robots}, explain code or a concept that they don't understand \cite{leinonen2023comparing}, and encourage better problem definition through prompt engineering \cite{denny2023conversing}.

One area in which very recent strides have been made is in using generative AI to explain cryptic programming error messages (PEMs) \cite{leinonen2023using}. Students and instructors have complained for the past six decades about how difficult it can be to understand some PEMs \cite{becker2019compiler}. Recent work has shown that enhancing PEMs via generative AI is very effective in actual classroom settings \cite{wang2024largescale, taylor2024dcc}. However, these have just been for compiler error messages.
 
In this paper, we explore the integration of generative AI into an automated assessment tool (AAT) in a CS1 course that provides feedback on compiler errors, run-time errors, and logic errors. We added real-time feedback from ChatGPT to one of the programming assignments in CS1 during the Fall 2023 semester and report on submission statistics as well as survey responses after the assignment was completed. Our hypotheses are that students will make fewer submissions with the guidance of ChatGPT and that students will find the real-time AI feedback helpful.

We are guided by the following research questions:

\begin{enumerate}
    \item[\textbf{RQ1:}] To what extent does real-time AI feedback impact student submission behavior when working on programming assignments? 
    
    \item[\textbf{RQ2:}] How do students perceive real-time AI feedback on their assignment submissions?
\end{enumerate}

The contributions of this work are:

\begin{enumerate}
    \item We implement generative AI into an automated assessment tool to enhance all programming error messages at both compile time and run time, as well as logic errors, for the first time.
    
    \item We discuss design considerations for integrating generative AI into automated assessment tools. The implications for \emph{how} students utilize such feedback can help both future researchers and tool creators make more usable interfaces.

\end{enumerate}

%
%
\section{Related Work}
Early work on LLMs in computing education centered on the capabilities of such tools. Already with GPT-3, which debuted in 2021, researchers found that generative AI was capable of solving common programming problems, such as the Rainfall Problem, with relative ease. Finnie-Ansley et al. found that it could solve most published variants of the problem with just the text description provided to the model. They also found that the model could answer exam questions, placing in the top quartile when compared to real student performance in a CS1 course \cite{finnieansley2022robots}. Follow-up work revealed similar performance in a CS2 course \cite{finnieansley2023my}. Recent work with GPT-4 has shown that current models outperform their predecessors by getting nearly every question correct, placing this model near the top spot when compared to humans \cite{prather2023robots, savelka2023thrilled}. 

Students are also using these tools to help them write code. Recent user studies on student behavior and interaction with LLMs, such as Github Copilot, have provided an interesting window into their use. Vaithilingam et al. looked at very early usage of students using LLMs to code and found that students preferred using it, despite it not decreasing their overall task completion times \cite{vaithilingam2022expectation}. The fact that it didn't make novices any faster at solving programming problems could be because students struggle to understand code that has been autogenerated for them, as found by Prather et al. \cite{prather2024tochi}. Kazemitabaar et al. found that students will utilize it to either explore new ways of doing a task or to attempt to accelerate their task completion time \cite{kazemitabaar2023studying}.

One way to help students is to use LLMs to produce code explanations. MacNeil et al. added code explanations to an ebook in a small study (n=30) and found that students rated them as understandable and helpful, though they did not engage with them as much as expected \cite{macneil2023experiences}. Leinonen et al. compared code explanations from students and GPT-3 in a large study (n=~1,000), rating the explanations for accuracy, understandability, and length. They found that LLM-created explanations were significantly easier to understand and more accurate than student-created explanations \cite{leinonen2023comparing}. 

Another related area of support that students have consistently needed over the past six decades is in understanding the feedback they receive about their programs \cite{becker2019compiler, prather2017novices, prather2018metacognitive,denny2021designing}. Leinonen et al. used Codex (a coding model built, at the time, on GPT-3) to interpret common PEMs and found the GPT-enhanced versions often surpassed the original PEMs in understability and actionability. Although showing promising results, the model they used is now outdated and they utilized a static approach. More recent work by Wang et al. implemented real-time GPT-4 feedback on compiler error messages in a large-scale course \cite{wang2024largescale}. They found that students repeated an error 23\% less often and resolved the error in 36\% fewer attempts, when compared to students using the original error messages. Taylor et al. implemented GPT feedback on compiler error messages into the Debugging C Compiler (DCC) in a very large course (n=2,565) \cite{taylor2024dcc}. They found that these messages were accurate 90\% of the time and were well received by students. However, they only provided feedback for compiler error messages. Given that there may be less of a need to provide feedback on syntax errors only -- because generative AI can write syntactically correct code -- it is important to now move beyond PEMs and into additional feedback on student submissions. 

Hellas et al. examined how GPT-3.5 would respond to novice programmer help requests \cite{hellas2023exploring}. They had previously collected a dataset of students who had pressed a ``help'' button and placed their help request into a queue for the instructor. When evaluating these requests via GPT-3.5, they noted that it often provided good feedback and almost always included code. In related work, Liffiton et al. created a tool for students (CodeHelp) to use where they could get automated help from an LLM for their coding problems. The tool allows them to paste the relevant part of their code, an error message, and a question into three separate boxes and then request help \cite{liffiton2023codehelp}. They found that students valued the tool's availability because it could be accessed when instructors or TAs were otherwise not available.

Our study extends the above work by using LLMs to providing feedback on all submissions, not just PEMs. We also extend the recent line of work on student help requests by having GPT responses to student submissions any time they submit and having that response provide feedback on their overall submission. Instead of having to ask for help or paste relevant bits of code and an error message into a tool, our implementation provides consistent help automatically.

\begin{figure*}
\centering
  \includegraphics[scale=0.5]{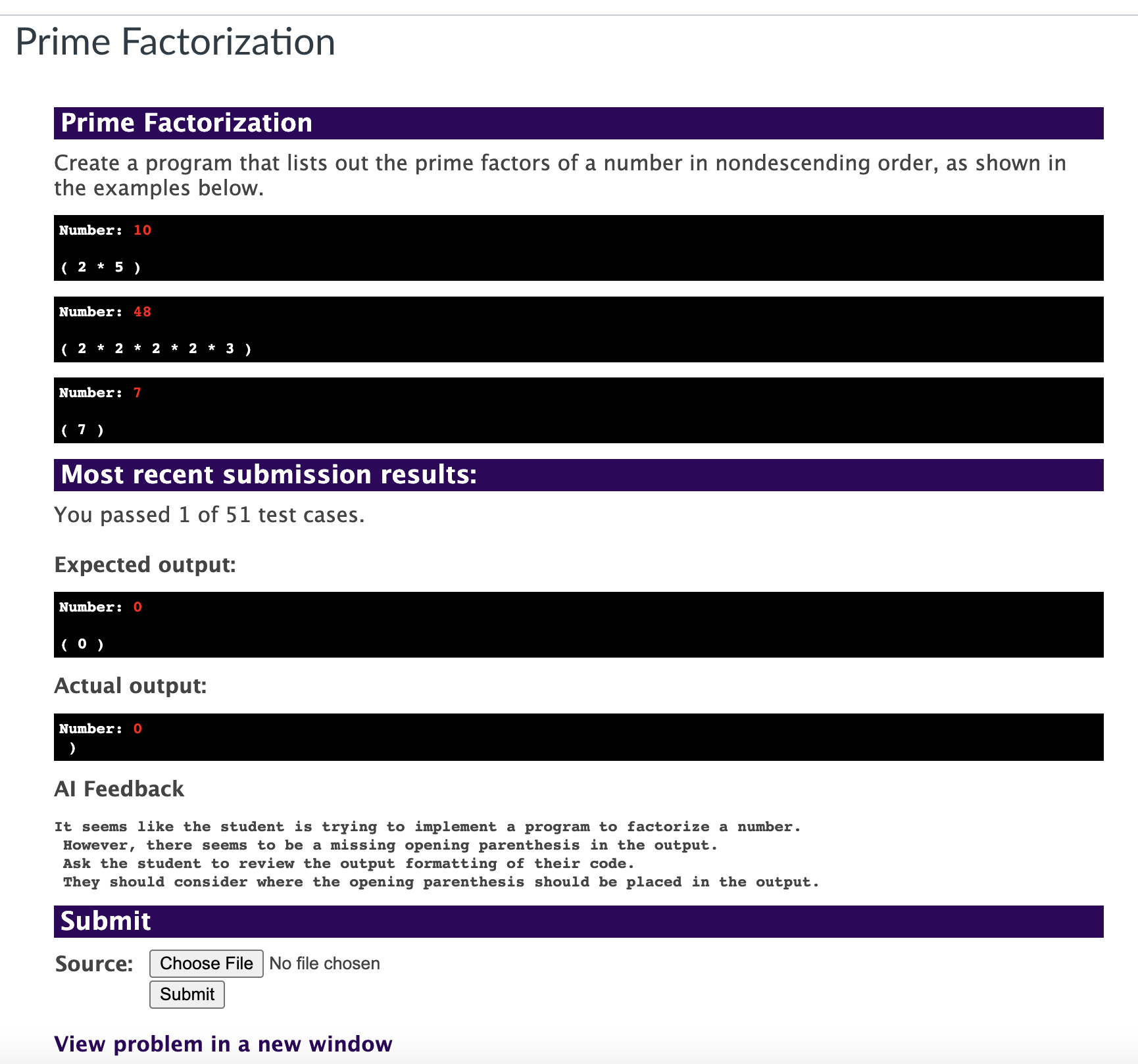}
  \caption{The problem description in \textsc{Athene} followed by the most recent submission results. This shows expected vs. actual outputs as well as the response from ChatGPT listed under "AI Feedback".}
  \label{fig:athene-gpt}
\end{figure*}

%
%
\section{Methodology}
Students in the introductory programming course at Abilene Christian University in the Fall 2023 semester (n=52) were recruited to participate in the study. All students were briefed on the study before participation by one of the authors and then provided consent forms. We also requested the previous ten years of submissions data from the Athene administrator. We gained approval for this study by the Abilene Christian University IRB committee. 

\subsection{Implementation}
We deployed a PHP plug-in to integrate with Athene, the automated assessment tool (AAT) used in our department. Athene has been used since 2009 at Abilene Christian University for providing automated feedback for CS1 programming problems \cite{towell2010walls}. Many studies have been done on enhancing the error messages served to students through Athene \cite{ pettit2017enhanced, prather2017novices, prather2018metacognitive, prather2019first}. This made our integration of ChatGPT with Athene a logical choice. 

A typical programming problem in Athene provides a text-based description of the problem and sample runs as examples for the student to see input transformed to output (see Figure \ref{fig:athene-gpt}). The student can then submit their code and receive feedback. If it does not compile, Athene returns the compiler error message. If it does compile, Athene will run the submitted code against test cases and return the number of test cases passed. Athene will return to the user an example of the first failed test case. Our plugin submits the student code to ChatGPT (running GPT-4) with a prompt before it. It was important that it not return code because we didn't want generative AI to solve the problem for the student. Given prior research on the difficulty of coaxing generative AI into providing a hint without also providing code \cite{hellas2023exploring, liffiton2023codehelp}, we crafted the following prompt over the course of many attempts:

\begin{quote}
    You are a programming tutor for an introductory programming course. You are supposed to help students without telling them what to do or how to do it. You cannot provide answers or straightforward instructions about how to fix their code. You are trying to help them learn themselves. Here is some code that a student wrote in this introductory programming course. Please provide a hint about what to do next, but do not tell the answer and do not provide any code in your feedback.\\
    Code:\\
    \textit{<code inserted starting here (in place of this comment)>}
\end{quote}

The plugin was integrated into one assignment, ``Prime Factorization''. Students were given this problem as an extra credit assignment and provided a week to complete it. Each time they submitted, they received feedback from ChatGPT under the label ``AI Feedback'' (see Figure \ref{fig:athene-gpt}). We collected all program submissions to Athene for this problem as well as survey data after the assignment due date had passed.

\subsection{Surveys}
Our first round survey was sent via email. We workshopped our questions with fellow students (who were not in the introductory programming course) as well as faculty. We chose Likert-style and yes/no questions in an attempt to collect more quantifiable data (see Tables \ref{tab:survey1-likert}, \ref{tab:survey1-yes-no}, \ref{tab:survey1-more-less}). Responses from the first round survey raised further issues around student perception of the GPT-enhanced feedback. We therefore sent a second round survey to collect additional data. This time, we chose to collect open response questions. These are also reported below in Section 4.2.

\subsection{Threats to Validity}
There are three primary threats to validity for our study. First, students could ask ChatGPT to solve their coding assignment before they turned it into Athene. There were no restrictions in place to prevent this from happening and our study was done in good faith that students would engage in the process as requested.
Second, students are encouraged to compile locally to check for errors before submitting to Athene for grading. 
Therefore, we do not capture all of their attempts. Rather, we only capture when they submit to Athene. 
Third, the professor of this course made this particular assignment for extra credit. This could skew participation rates.

%
%
\section{Results and Discussion}
\begin{figure}
\centering
  \includegraphics[scale=0.46]{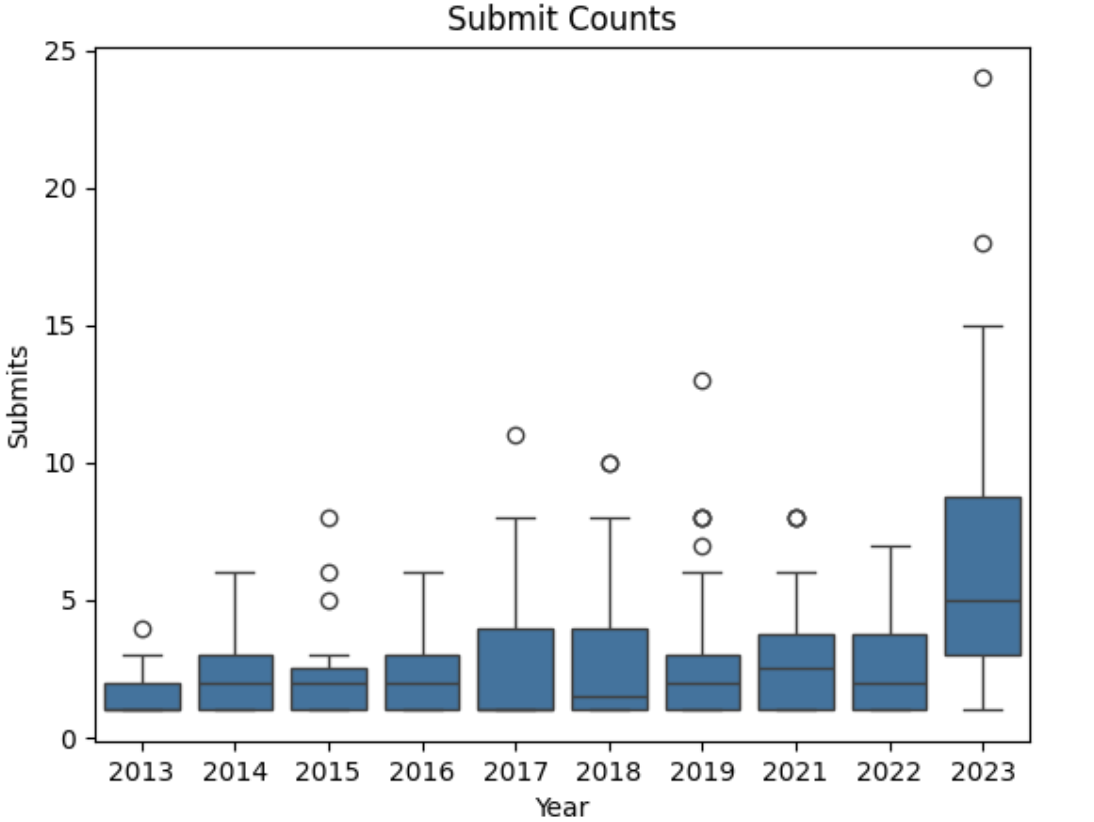}
  \caption{Submission data visualized from the experiment (2023) and the previous ten years for the same problem, Prime Factorization.}
  \label{fig:submissions-plot}
\end{figure}

We examined the submission logs for the Prime Factorization problem during the Fall 2023 semester as well as the previous ten years for the same problem in Fall semesters. Even though 50 students signed the consent forms, only 42 attempted the Prime Factorization programming problem. The mean number of submissions in fall 2023 was 6.405 (SD = 5.133, min = 1, max = 24). This is quite a bit higher than in previous years (see Figure \ref{fig:submissions-plot}). Our hypothesis was that there would be fewer submissions on average in 2023, which was proved false. A look at the means reveal that 2023 has a significantly higher mean submission rate than the next highest year (2018), and consequently, all other years (t(97) = 3.60, p < .05). This would be in direct contrast to recent work that showed real-time AI feedback lowered the number of submissions \cite{wang2024largescale,taylor2024dcc}. However, we did not measure learning outcomes, so there could be other explanations for the increase in submissions. Therefore, we turned to the survey data to help find an explanation for this behavior.


\subsection{First Round Survey}
The first round was sent immediately after the assignment closed (one week after opening) and included 12 questions. Thirty-three students responded to the first round survey. We anticipated that students would find the real-time AI feedback helpful. However, we received a mixed set of opinions. With regard to the Likert-scale questions, all averages fell to around 3.0 (see Table \ref{tab:survey1-likert}). The yes/no questions (see Table \ref{tab:survey1-yes-no}) also revealed a similarly mixed response to the AI feedback, with the notable exception of the question, \emph{Would you like the AI assistance on other assignments in the future?}, which received 69.7\% of students saying ``yes''. Most students indicated that having the AI feedback did not help them learn or that having AI feedback made no difference (see Table \ref{tab:survey1-more-less}). It seemed students were not too impressed with AI feedback, but were still interested in using it again. Therefore, we sent a second round survey with open-ended responses in an attempt understand the mixed responses and the higher submission rate.

%
%
%
%
\begin{table*}
\renewcommand{\arraystretch}{1.5}
\begin{tabular}{p{0.9\textwidth}|c}
\textbf{Question (Scale: 1-Strongly Disagree -> 5-Strongly Agree)} & \textbf{Mean} \\
\toprule

Having AI assistance helped me complete the assignment easier.  & 2.87 \\

\hline
The AI feedback messages provided by ChatGPT were more helpful than feedback in previous assignments without AI.  & 3.33 \\

\hline
AI helped prevent me from seeking any other external resources for this assignment (stack overflow, tutoring...etc).  & 2.87 \\

\hline
If I always had AI feedback in Athene, I would use less outside resources.  & 3.12 \\

\hline
I was frustrated with Athene in previous assignments this semester.  & 3.09 \\

\hline
I was frustrated with Athene in the Prime Factorization problem that utilized AI feedback.   & 2.96 \\

\bottomrule
\end{tabular}
\caption{Response data from the first survey on Likert-style questions.}
\label{tab:survey1-likert}
\end{table*}

%
%
%
%
\begin{table*}
\renewcommand{\arraystretch}{1.5}
\begin{tabular}{p{0.85\textwidth}|cc}
\textbf{Question (Yes or No response)} & \textbf{Yes} & \textbf{No}\\
\toprule

Do you think this assignment on average took you less time than normal?   & 7 & 26 \\
\hline
Do you think the AI assistance helped you actually understand what you were coding more effectively?   & 18 & 15 \\
\hline
Would you like AI assistance on other assignments in the future?    & 23 & 10 \\
\hline
Did using AI in this assignment make you want to pursue coding as a career more?    & 12 & 21 \\
\hline
Did this assignment make you more confident in coding?    & 14 & 19 \\
\bottomrule
\end{tabular}
\caption{Response data from the first survey on yes/no questions.}
\label{tab:survey1-yes-no}
\end{table*}

%
%
%
%
\begin{table*}
\begin{tabular}{p{0.67\textwidth}|ccc}
\textbf{Question (More/No Difference/Less)} & \textbf{More} & \textbf{No Difference} & \textbf{Less} \\
\toprule

Do you think you learned more or less during this assignment? & 14 & 16 & 3 \\

\bottomrule
\end{tabular}
\caption{Response data from the first survey on our final question with responses yes/no difference/less.}
\label{tab:survey1-more-less}
\end{table*}



\subsection{Second Round Survey}
The second round survey consisted of four questions and was sent out a few days after the first. The questions were:

\begin{enumerate}
    \item Please tell us about your experience using the AI feedback in the assignment.
    \item Did you like or dislike AI feedback on your programming assignment? Please explain why.
    \item What do you think could have made the AI feedback more helpful for you?
    \item Was the AI feedback ever wrong? If so, please describe what happened and how you dealt with it.
\end{enumerate}

We examined all nine responses to the four questions and discussed the most commonly occurring themes together.

\subsubsection{Vague or Incorrect Advice:} The most commonly mentioned topic in the survey was that the AI feedback was often too vague. Eight of the nine respondents mentioned this, sometimes multiple times. P4 wrote, \emph{``I felt it was pretty vague at times. At one point it said that I should check my loop. The problem with that was I had two loops at the time, and it did not specify which one.''} And P5 wrote, ``It wasn't wrong because it never really provided any useful solution.'' These quotes illustrate student frustration with AI feedback that never quite helps them enough and could explain the lukewarm responses to the first survey. P1 also wrote, \emph{``It gave me feedback to change a line of code that produced a compiling error''}, which could also have contributed to the lack of reception of the tool in the first survey. The fact that AI could provide incorrect feedback has been noted in the literature and is concerning enough for some to seek to ban the use of generative AI tools altogether \cite{lau2023ban}. One solution utilized in the CodeHelp tool by Liffiton et al. was to prominently place a warning above the feedback that it could be incorrect \cite{liffiton2023codehelp}. 

\subsubsection{Helpful:} Four of the nine respondents wrote that they found the AI feedback helpful. For instance, P6 wrote, \emph{``It made the coding process significantly easier''}. So, it seems that some students were helped more than others. However, this positive perception could be caused by students receiving more help than usual. For instance, P2 wrote, \emph{``I liked it because it gave more thorough feedback than regular Athene''}. Usually Athene only provides programming error messages or test case failure data, so some students felt that any additional information was an improvement. This finding is in line with user studies on code explanations \cite{macneil2023experiences, liffiton2023codehelp}.

\subsubsection{Interaction:} Five of the nine respondents mentioned the interaction style of the AI feedback. P2 and P3 wrote that it should provide narrower answers, instead of the high-level broad feedback we directed it to provide through our prompt. P5 wrote that, \emph{``I did not like the AI feedback because it was very generic and there was no way to converse with it. It only gave me feedback when I submitted code and I could not add any other commentary.''} And P6 wrote, \emph{``Enabling it to learn from previous submission.''} Students wanted to interact with the AI, rather than see a single-shot hint or bit of advice. It seems that students want the ability to engage the AI system in a conversation about their error, asking clarifying questions, or to simply allow it to use a conversation thread to provide better answers in context. However, as Liffiton et al. note, using a one-shot approach without the possibility of follow-up or dialogue (like the one in this experiment) makes it less likely that the model goes ``off the rails'' or produces harmful or biased responses \cite{liffiton2023codehelp}.

One student wrote about what drove their engagement with the system. P7 wrote, \emph{``It was quite simple to use AI feedback for this assignment because you could keep submitting your code and verify what it is that you're missing or not getting through.''} This could be the answer to why students submitted so many more times on average during the experiment semester and why a majority wanted AI feedback in future assignments. It's possible that students submitted more often because they hoped to receive advanced feedback when they were stuck, whereas students in previous semesters wouldn't have bothered until their local output matched whatever failed test case Athene had given them. A similar surge in engagement was noted by Liffiton et al. \cite{liffiton2023codehelp} in their CodeHelp tool. Implementing AI feedback has the potential to shift the student experience. Instead of dreading to submit for fear of receiving the same response, the student is encouraged to experiment and receive new feedback each time.

%
%
\section{Conclusion}
In this paper, we presented a study on adding generative AI feedback to an AAT and tracked student code submissions and perceptions of the experience. From all the data presented above, it seems that merely adding generative AI feedback to an AAT does not necessarily make it better. Our data indicates that the interaction style could impact the tool's usefulness and trustworthiness. Students want more feedback and help on their programs, and are willing to engage with AI feedback, but are wary of it being too vague or even incorrect and therefore a waste of their time. Similarly, they expect to have the normal affordances \cite{macneil2023prompt} that have become the gold standard of generative AI interaction, such as chat-type interfaces where they can ask follow-up questions. Future work should explore designing an interface that allows limited in-context follow-up to AI submission feedback.

%

\begin{acks}
Thank you to Dr. Solofoarisina Arisoa Randrianasolo for allowing us to utilize your CS1 course for this experiment. Thank you to Dr. Dwayne Towell for helping us collect the data from Athene. Thank you to Dr. Brent Reeves for helping us analyze the collected data. Thank you to Dr. James Prather for mentoring the ACU SIGCHI Local Chapter. Finally, thank you to Dr. Paul Denny, Dr. Brett Becker, and Dr. Juho Leinonen for reviewing this paper before submission.
\end{acks}

%
\balance

%
\bibliographystyle{ACM-Reference-Format}
\bibliography{references}

\end{document}